# SENDING AND SEARCHING FOR INTERSTELLAR MESSAGES

Alexander L. Zaitsev

[Institute of Radio Engineering and Electronics](), Russia

*…there is nothing more dangerous than to speak about the danger of communication.*
Vitaly L. Ginzburg, Nobel Laureate, [1, p. 348].

## ABSTRACT

There is a close interrelation between Searching for Extraterrestrial Intelligence (SETI) and Messaging to Extraterrestrial Intelligence (METI). For example, the answers to the questions "Where to search" and "Where to send" are equivalent, in that both require an identical selection from the same target star lists. Similar considerations lead to a strategy of time synchronization between sending and searching. Both SETI and METI use large reflectors. The concept of "magic frequencies" may be applicable to both SETI and METI. Efforts to understand an alien civilization's Interstellar Messages (IMs), and efforts to compose our own IMs so they will be easily understood by unfamiliar Extraterrestrials, are mutually complementary. Furthermore, the METI-question: "How can we benefit from sending IMs, if a response may come only thousands of years later?" begs an equivalent SETI-question: "How can we benefit from searching, if it is impossible now to perceive the motivations and feelings of those who may have sent messages in the distant past?" A joint consideration of the theoretical and the practical aspects of both sending and searching for IMs, in the framework of a unified, disciplined scientific approach, can be quite fruitful. We seek to resolve the cultural disconnect between those who advocate sending interstellar messages, and others who anathematize those who would transmit.

## INTRODUCTION

Information interaction of Cosmic Civilizations means both reception and sending of interstellar messages (IMs). Search of IMs has sense only in the case when those who search assume existence of IM sources. Similarly, sending IMs is meaningful only if there is a hope for existence of those who can detect these IMs. Thus, the concept "Interstellar Messages" applicable to both SETI (treated as searches of IMs) and METI (treated as sending of IMs).

The given concept can be considered as a specific invariant of transformation SETI ⇔ METI. As such an approach there is no division into Terrestrial and Extra-Terrestrials and opposition of the Earth and Cosmos that allows considering attempts of informa-tion interaction of intelligence space systems as something universal, inherent to the inhabited Universe. Besides, it is important to note that at such an approach one does not talk any longer about "Messages from the Earth" but about Interstellar Messages. This may lower the tension in the discussions with those who feel superstitious fear and anxiety as soon as sending and transmissions are mentioned.

At the searches, two aspects are analyzed. First (1), how to answer such questions related to Searches of IMs as "Where to search?", "Whether there is a sense to search?", "Whether searches are dangerous?", etc., based on the current science about the nature and a society and technological level. Second (2), how the Sender acted from the point of view of conducting searches and what purposes it pursued sending IMs ("Model of the Cosmic Civilization who sends IMs"). Two similar aspects can be analyzed at sending: (3) how to answer such questions of IM Sending as: "Where to send?", "Whether there is a sense to send?", "Whether sending is dangerous?", etc., based on the current state of science about the nature and society, and technological level, and (4) how the addressee will act from the point of view of conducting sending and what actions it will undertake at the detection of IMs" ("Model of the Cosmic Civilization conducting searches of IMs"). Altogether, we come to *four aspects* of the problem of sending and searching for IMs.

Below the concept of *four aspects* is explained using the above-mentioned approach to SETI and METI [2], but formulated from the uniform position of sending and searching for IMs and in terms of the information interaction of Cosmic Civilizations:

1. In what is the sense of sending and searching for interstellar messages?
2. Where to send and where to search for IMs?
3. The dangers related to sending and searching for IMs.

## WHAT IS THE MAIN IDEA BEHIND SENDING AND SEARCHING FOR INTERSTELLAR MESSAGES?

It is considered, that the main idea of searches is obvious and trivial; it consists in an opportunity to receive valuable information. But it is not as simple as seems. Really, how can we benefit from searching, if it is impossible now to perceive the motivations and feelings of those who may have sent IMs in the distant past? What for to send IMs? And whether there is basically such need, as sending IMs? If we declare that we can explain our reasons for searches and we can prove the need of sending signals, then the search gets meaning as proved by the existence of the subject of the search. It has been already noted repeatedly [see, e.g. 3] that sending and searches are in close indissoluble interrelations. Only after we understand (or, on the contrary, after we have not managed to understand) what for we need sending IMs and if such an unselfish and messianic activity is natural for a developed civilization, we can prove the searches themselves as well as that SETI is meaningful (or, on the contrary, it is not meaningful). Also, after we understand what is the need for Intellect to send information to prospective Others, we naturally come to understanding the sense of our own transmissions. So far a question: "Is sending IMs some indispensable attribute of Intellect?" is not answered. And, hence, it is not answered the question: "Does SETI have sense?"

Sure, we leave aside two such exotic explanations, as: (1) "Sense of SETI consists in searches not purposeful transmissions, but leakage of electromagnetic radiation" and (2) "Sense of sending is a kind of "fishing" by aggressive super-civilizations of trustful and ingenuous, naive and unripe civilizations such as the terrestrial one." If we accept them, then SETI will get a simple role of a tool to search for such potential "Star Aggressors" and "Star Interventionists" and with the unique purpose to find them and hide without any response to them.

## WHERE TO SEND AND WHERE TO SEARCH FOR INTERSTELLAR MESSAGES?

In addition to traditional criteria of target star selection [4] for both SETI and METI, there is a number of additional questions under joint "Sending & Searching" consideration. For example, does our star fits as a candidate for sending IMs? Or, is there a hope (and if yes, what it is based on) that Others will choose the Sun as the addressee of IMs and will put our star on the target-list? Similarly regarding sending IMs: are we objects of search for those whom we choose as addressees of own sending? May our efforts be worthless since from Their point of view, our Sun does not represent absolutely any interest as the object of search? And so on…

## THE DANGER RELATED TO SENDING AND SEARCHING FOR INTERSTELLAR MESSAGES

Quite often one can hear cautions to those who under own initiative, without a sanction of the United Nations or a similar international organization, sends IMs. The argument of opponents of sending initiative IMs is well-known, it can be found, for example, in [5] and there is no need here once again to repeat it. However, to be consistent, it is necessary to agree that uncontrolled searches are also unsafe. If a country receives a certain "premature knowledge" as a result of a search not controlled by the United Nations or a similar international organization and this country is not ready from the moral-ethical point of view, this country (or a coalition of the countries) may use it to harm the rest of the mankind. Imagine that some morally ugly creature or a religious fanatic with the ideas on the level of the Stone Age suddenly receives the secrets of a terrible and powerful weapon! Thus, it is necessary to keep SETI under some effective international control. In other words, in case of using the concept "Sending and searching for IMs", the shift from a specific question "Is METI dangerous?" to more general question "Is such human activity as sending and searching for intelligence signals in the Universe dangerous in principle?" is quite reasonable.

As to the danger related to transmission of interstellar radio messages (IRMs), a more careful analysis shows that the pointed radiation of IRMs sent using planetary and asteroid radar telescopes, apparently, is not as dangerous as pointless transmissions of the same radars. Fig. 1 indicates all transmissions during radar observations of planets, asteroids and comets. Data on radar observations of asteroids and comets, carried out by the radars located in Arecibo (347 red points), Goldstone (661 blue points), and Evpatoria (215 green points), are taken from [6], and on a radar observations of planets (unfortunately, only for Goldstone and Evpatoria) – from [7, 8].

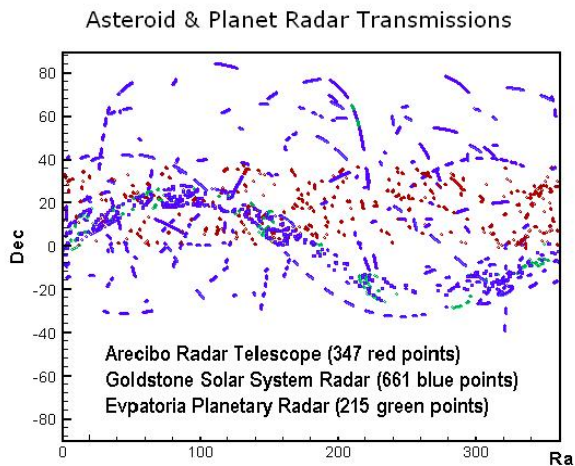

The analysis of radar data has revealed the following experimental fact: any among the 1223 transmissions does not get to stars [9]. This means that the interstellar space is almost empty; the distance between stars is much greater than the size of stars and "belts of a life" around the stars. Therefore at pointless casual radiation transmission the probability of getting into inhabited zones is insignificantly small. It is important to mention the following feature of the radar observations of Solar system bodies: a slow scanning over the celestial sphere that is related to the proper motion of targets of the radar observation. From this fact two important conclusions follow. First, this may explain why we do not detect any radar signals from other civilizations. Ostro and Sagan [10] explained the absence of signals from Their radar telescopes by the idea that They may not use a radar astronomy and, consequently, are not protected against the asteroid or comet hazard. We have another, rather reasonable explanation. If the probability of our radar transmissions to get into the habitable zones of cataloged stars is very low and They do not see us, then the probability to get to the Earth at similar pointless transmissions implemented by other civilizations is also very low. For this reason, we also do not see Them.

The second, not less essential conclusion from that fact that any of our 1223 transmissions have not get into a habitable zone of the Type I civilizations (the civilizations of the type higher than the first one live "practically everywhere", not just near the parent stars, [11]) when the radar beams slowly scan the sky, illuminating greater areas of the Galaxy, consists that such radiation is much easier to be detected by those unknown aggressive and super-power civilizations which scare so much the METI-opponents. In this sense rare pointing transmissions of interstellar radio messages represent considerably smaller danger than numerous addressless radar astronomy transmissions. There are two reasons for that. First, IRMs are precisely directed to target specific stars, and, second, the radar beam is motionless relatively to other stars and, hence, during radiation does not pro-vide any scanning and does not illuminate gradually celestial sphere (Fig. 2).

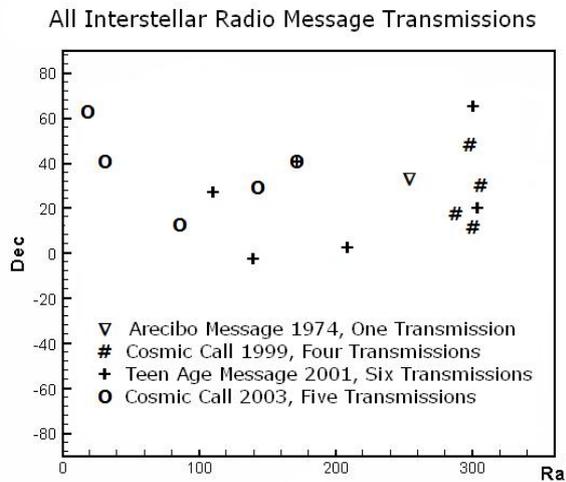

Thus, in order to be detected by some young Type I civilizations living nearby the parent stars, it is necessary to address our IRM transmissions. Accidental detection by such civilizations of signals from the planetary and asteroid radars of some Other civilization is extremely unlikely. If we are afraid of powerful and aggressive civilizations of Type II and Type III, which live "practically everywhere", it is necessary to forbid numerous pointless transmissions of asteroid and planetary radars as their radiation gradually illuminates greater areas that promotes its detection by "star aggressors and interventionists".

However, it is clear that a ban on radar investigations of small solar-system bodies makes it impossible to provide a protection against asteroid and comet hazard [12]. Moreover, we can see a rapid growth in the number of new radar detections of asteroids and comets [13], Fig. 3, and this tendency will grove even stronger when more powerful and dedicated asteroid and comet radar systems will be created, [14]. This will result in more complete coverage of the celestial sphere by terrestrial power electromagnetic radiation.

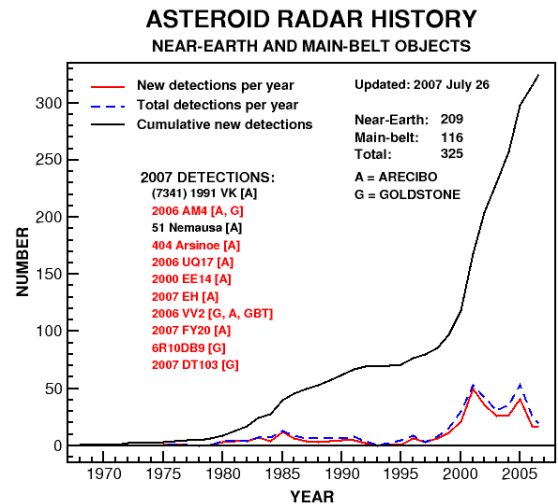

Thus, the notorious thesis that it is pointed radiations of IRMs that represent the fatal *"danger"* should be ruled out from the agenda. Therefore, we guess that it is quite reasonable now to try to use Arecibo Radar Telescope and Goldstone Solar System Radar, along with the Evpatoria Planetary Radar, which was already used for Cosmic Call 1999 & 2003 and Teen Age Message 2001 transmissions [15], as interstellar radio message transmitters. These radars have a few times greater energy potential than the Evpatoria one [16], so they can provide more efficient sending of further IRMs.

*Acknowledgments.* I thank Allen Tough for valuable remarks and financial support.
I am grateful to Ludmilla Kolokolova for her useful comments on the manuscript.

## REFERENCES

1. Communication with Extraterrestrial Intelligence (CETI), by Carl Sagan (Editor). Cambridge & London: The M.I.T. Press, 1973, http://worldcat.org/isbn/9780262690379

2. A. Zaitsev. Messaging to Extra-Terrestrial Intelligence, http://arxiv.org/abs/physics/0610031

3. A. Zaitsev. The SETI Paradox, http://arxiv.org/abs/physics/0611283

4. M. Turnbull and J. Tarter. Target Selection for SETI: 1. A Catalog of Nearby Habitable Stellar Systems, http://arxiv.org/abs/astro-ph/0210675

5. D. Brin. Shall We Shout Into the Cosmos? http://www.davidbrin.com/setisearch.html

6. JPL Solar System Dynamics, http://ssd.jpl.nasa.gov/?radar

7. Planetary Ephemeris Data, http://iau-comm4.jpl.nasa.gov/plan-eph-data/index.html#radar

8. Russian Radar Ranging of Planets (1962-1995), http://www.ipa.nw.ru/PAGE/DEPFUND/LEA/ENG/rrr.html

9. D. Churakov. E-mail communication on June 28, 2007.

10. S. Ostro and C. Sagan. Cosmic Collisions and Galactic Civilizations. Astronomy & Geophysics: The Journal of the Royal Astronomical Society, 39 (4), 1998, 22-24. See also: http://trs-new.jpl.nasa.gov/dspace/bitstream/2014/19498/1/98-0908.pdf

11. Kardashev civilizations, http://www.daviddarling.info/encyclopedia/K/Kardashevciv.html

12. Near-Earth Object Survey and Deflection Analysis of Alternatives, NASA Report to Congress, March 2007, http://www.nasa.gov/pdf/171331main_NEO_report_march07.pdf

13. Asteroid Radar History, http://echo.jpl.nasa.gov/asteroids/PDS.asteroid.radar.history.html

14. S. Ostro, A. Zaitsev, Y. Koyama, and A. Harris. Dedicated Asteroid and Comet Radar. XXIII-rd IAU GA Abstract Book, 1997, Kyoto, Japan, 50, http://fire.relarn.ru/126/docs/iau23kyoto.pdf

15. Interstellar Radio Message (IRM), http://www.daviddarling.info/encyclopedia/I/IRM.html

16. A. Zaitsev. Limitations on Volume of Interstellar Radio Messages, http://www.cplire.ru/html/ra&sr/irm/limitations.html